\documentclass{kapproc} 

\usepackage{procps} 

\usepackage[dvips]{graphicx}

\upperandlowercase

\kluwerbib 

\begin{document}

\articletitle[The Dusty-Gas Disk of NGC\,3656]
{The Dusty Disk of the Early Galaxy NGC\,3656} 



\author{Lerothodi L. Leeuw,\altaffilmark{1}
Jacqueline Davidson,\altaffilmark{2} C. Darren Dowell,\altaffilmark{3} 
Roger H. Hildebrand,\altaffilmark{1} and Henry E. Matthews\altaffilmark{4}}


\affil{\altaffilmark{1}University of Chicago (USA), \ 
\altaffilmark{2}USRA-SOFIA (USA), \ 
\altaffilmark{3}Caltech-JPL (USA), \ \&
\altaffilmark{4}NRC-HIA (Canada)}

\begin{abstract}
SHARC\,II 350\,$\mu$m continuum and archival $HST$ $J-H$ band maps
are presented of NGC\,3656, the brightest of our sample of six elliptical galaxies for which
resolved CO gas disks have recently
been detected with $7''$-spatial-resolution, interferometry mapping.  
These gas disks confirm the conclusions of
earlier results showing optical dust lanes and unresolved CO that implied
the common existence of molecular gas in ellipticals 
and the disk-like structure of this gas.
The presented SHARC\,II mapping results
provide the 
best to date 
resolved
FIR-submm extent of NGC\,3656 and of any elliptical galaxy $>40$\,Mpc, 
showing that dust of 29\,K
exists out to at least $\sim 1.8$\,kpc in this galaxy.
These new data are used in conjunction
with the archival $HST$ maps and other published data 
to determine 
dust properties and associations with galactic structures, including
dominant heating 
sources such as 
star-formation or diffuse-stellar radiation. 


\end{abstract}


\section{Introduction}
Cold gas and dust in nearby elliptical galaxies was discovered only
about 15 years ago. Compared to the cold ISM in spirals, the cold gas
and dust is present in relatively small amounts and is seen in only
50\% to 80\% of nearby ellipticals.  The source and content of the
cold ISM in these galaxies is still uncertain, with optical and
far-infrared dust-mass estimates differing by $\sim$ 10 to 100.  Therefore,
the present study aims to map and analyze emission from the cold dust
of elliptical galaxies in order to probe and better
constrain the dust and evolutionary properties of ellipticals.  

We begin with a small sample of far-infrared (FIR) bright, nearby
(less than 70\,Mpc) ellipticals for which gas disks of CO emission have
recently been resolved with interferometry of  7$''$ spatial
resolution (e.g. Young 2002; see {\sl left} panel of
Figure~\ref{fig:leeuw1}). Our observational study exploits the
latest very sensitive submm array detectors 
(e.g., SHARC\,II at 350\,$\mu$m on the CSO) with a goal
of providing the best submm (cold dust) distribution maps to date for
our sample of galaxies (and for most in the sample the very first
detections beyond 100\,$\mu$m); dust temperature maps; gas-to-dust mass
ratios; dust grain properties; dust association with other galactic
structures; and constraints for models of dust evolution and
generation in elliptical galaxies.  


\section{
NGC\,3656 and other Elliptical Galaxies in the Sample}

The six galaxies in our 
sample are all ellipticals in that their
luminosity profiles follow the de Vaucouleur law; however, they
represent a spread of merger traces or ages, from galaxies that have
been classified as on-going or early-age major merger (e.g. NGC\,3656,
Balcells et al. 2001) to very-late accretion or quiescent system
(e.g. NGC\,807, Murray
et al. 2000).
The FIR and submm mapping results (e.g. 
dust content) will
be compared between the sample 
galaxies as a probe of not only dust but also of local
merger-formation and evolutionary history of elliptical galaxies in
general. 

\begin{figure}[!ht]
\vspace{-0.4cm}
\begin{center} 
\includegraphics[angle=0,width=2.25in]{leeuw_l_fig1l.ps}
\includegraphics[angle=0,width=2.25in]{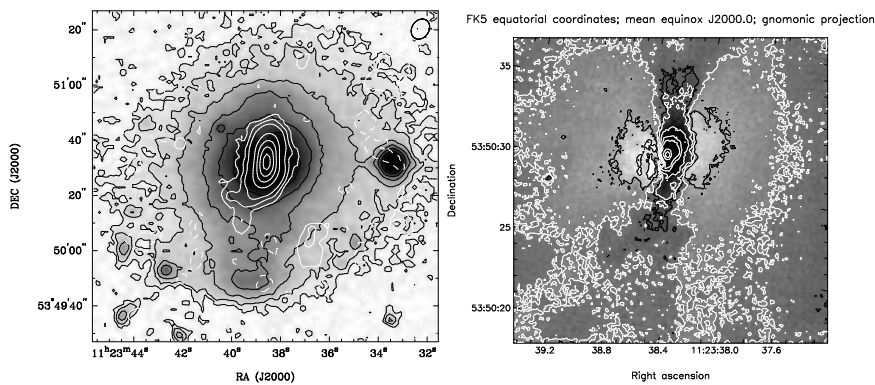}
\vspace*{-0.3cm}
\caption[]{
\small{
{\sl Left:}
Total integrated CO(1-0) intensity map (white
contours) of NGC\,3656
with CO diameter of $34''$
overlaid on optical
data (gray scale and black contours) 
adopted from 
by Young 2002.  
The white contours are in units of -5\%, -2\%, 2\%, 5\%, 10\%, 20\%,
30\%, 50\%, 70\%, and 90\% of 
81.1\,Jy $\rm{beam}^{-1} \rm{km} \rm{s}^{-1} = 4.7 \times 
10^{22} \rm{cm}^{-2} $
CO integrated intensity or column densities peaks
(Young 2002). 
{\sl Right:} 
An archival $HST$ FW110 image overlayed with FW110-FW160 
contours to depict the dust extinction in NGC\,3656.
}
}
\label{fig:leeuw1}
\end{center}
\vspace*{-0.8cm}
\end{figure}

NGC\,3656 is the far-infrared (FIR)-brightest elliptical in our
sample. It has an optical elliptical body with an obscuring
north-south, galactic minor-axis, edge-on gaseous dust lane (see {\sl right} panel of
Figure~\ref{fig:leeuw1}), two tidal tails, a
system of shells and counter-rotating cores.  These features together
have been interpreted as evidence that the NGC\,3656 system is an early
major-merger remnant of disk galaxies (c.f. Balcells 1997, 2001).
Central structures that are seen in unsharp-masked and residual
galaxy-model K-band images of NGC\,3656 have recently been interpreted
as qualitative evidence that phase-mixing, since the disk-disk merger
and subsequent violent relaxation of this galaxy, is incomplete
(Rothberg \& Joseph  2004).

\section{
Archival $HST$ and CSO observations of NGC\,3656}

The {\sl right} panel in
Figure~\ref{fig:leeuw1}
shows an $HST$ FW110 image with FW110-FW160 ($\sim J-H$ band) contours
that demonstrate that the dust lane
of NGC\,3656 has the most extinction in a north-south region 
of radius $\sim 5''$ with
east-west asymmetry and a peak that is centered about 1$''$ east
of the galactic nucleus (or the center of the dust lane).  The asymmetry is
consistent with the dust lane being seen edge-on and its nearside
being on the galaxy's eastern part.   

\begin{figure}[!hbt]
\vspace{-0.3cm}
\begin{center}
\includegraphics[angle=0,width=1.8in]{leeuw_l_fig2.eps}
\includegraphics[angle=0,width=2.3in]{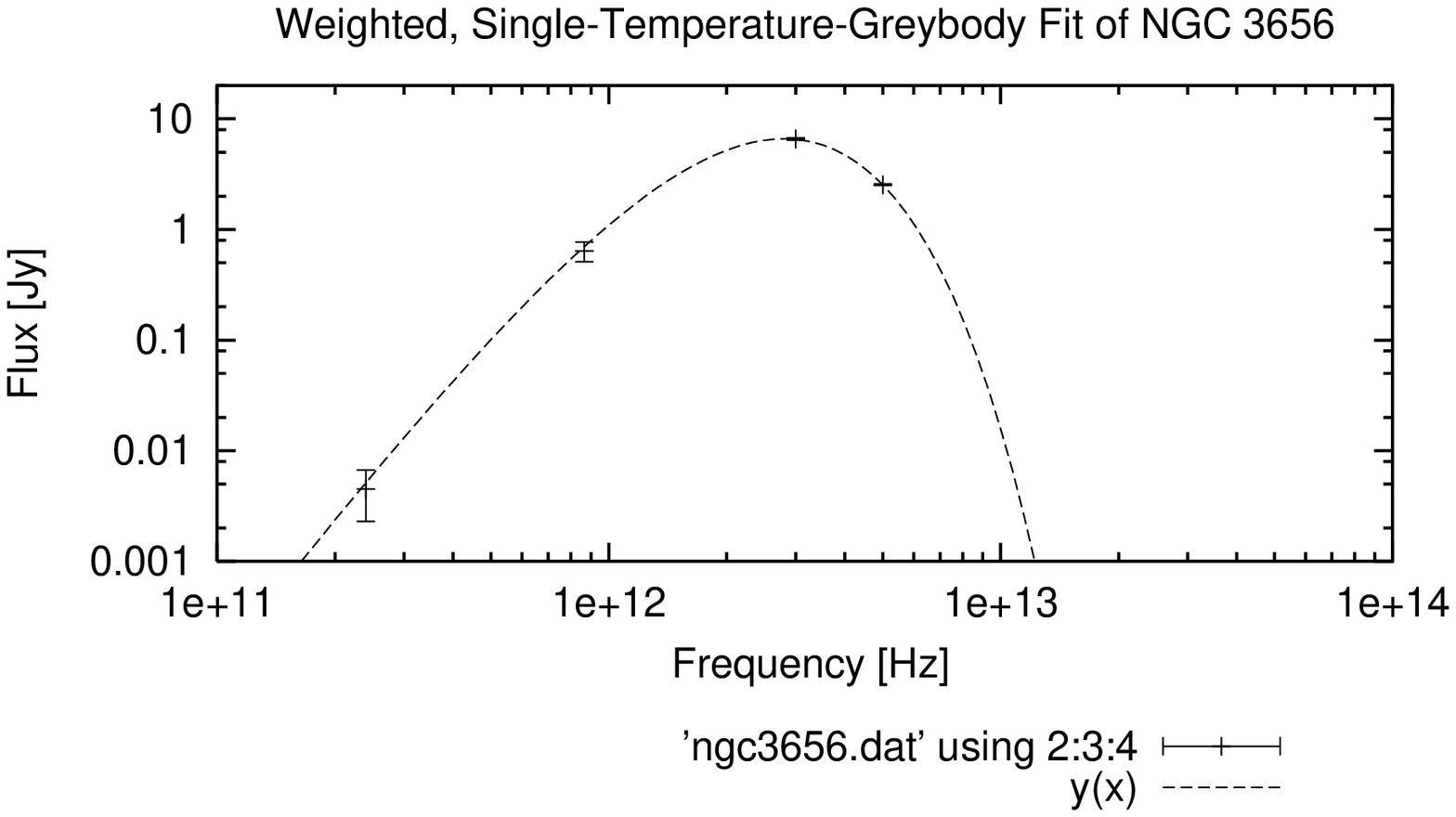}
\sidebyside
{ 
\vspace*{-0.7cm}
\caption[]{
\small{
SHARC\,II 350\,$\mu$m continuum map of NGC\,3656 
obtained with a CSO beam of $9''$.  The submm contours are in units of
0.06, 0.1, 0.14, 0.18, 0.22, 0.26, 0.3, 0.34, 0.38\,Jy
$\rm{beam}^{-1}$.
}
}
\label{fig:leeuw2}\vspace*{-0.8cm}}
{
\vspace*{-0.8cm}
\caption[]{
\small{
Weighted, single-temperature greybody of 29\,K and emissivity
index of 1.6
fit to $IRAS$ 60\,$\mu$m and
100\,$\mu$m, SHARC\,II 350\,$\mu$m and IRAM 1250\,$\mu$m data of NGC\,3656.
}
}
\label{fig:leeuw3}
}
\end{center}
\vspace*{-0.8cm}
\end{figure}

Figure~\ref{fig:leeuw2} shows a  map of NGC\,3656
from our exploratory
CSO/SHARC\,II observations, 
during which 
we detected
spatially-resolved, 350\,$\mu$m continuum 
emission. 
The submm continuum of
NGC\,3656 has an unresolved core and extended emission that is
slightly 
elongated north-to-southly down to the
50\% contour level and within a radius of $\sim 5''$.
Beyond this it is detected with less certainty and has a generally
bulging S-shape in the similar sense as 
seen in 
CO (1-0) and H-alpha
images of the same region (see, e.g., 
{\sl left} panel in 
Figure~\ref{fig:leeuw1}).

The extent of the resolved 350\,$\mu$m 
emission down to the
50\% contour level in NGC\,3656 is
consistent with a de-convolved Gaussian of FWHM of 8$''$.1.
This is about half the extent of the CO (1-0) emission observed in
this galaxy.  If one assumes the 350\,$\mu$m emission is associated
with the same dust as 
measured by $IRAS$ at 60 and 100\,$\mu$m, then
our CSO results imply 29\,K dust with an emissivity
index of 1.6 and angular size of $1.5 \times 10^{-9}$\,steradians (see
Figure~\ref{fig:leeuw3}).  Using 
the distance of 45\,Mpc (e.g. Young 2002) and 350\,$\mu$m dust
absorption value of 0.192 
m$^2/$kg (Draine 2003), a dust mass of $1.4 \times 10^8$ solar masses is 
calculated for the temperature of 29\,K and 350\,$\mu$m  integrated
flux of 0.64\,Jy.

\section{Implications of the CSO and $HST$ results 
of NGC\,3656}

Both the flux level and extent of the submm emission show that our
350\,$\mu$m  map of NGC\,3656 cannot represent all the dust associated with
the CO gas as resolved in this galaxy by Young 2002, if the normal
gas-to-dust ratio (100) is used to estimate the dust flux level
expected at 350\,$\mu$m.   The high 
extinction region
of radius $\sim 5''$ seen in the 
$HST$ colour map is
roughly co-spatial with the high-brightness 
submm and radio regions respectively presented in this paper and by
Mollenhoff et al. 1992, and probably represents a
star-formation site and thus warmer and/or denser dust in NGC\,3656.
Our SHARC\,II 
data constrain the temperature 
of the remainder, more extended, dust distribution (associated with
the more extended CO emission) to be less than 20\,K.   
The current submm observations
do not resolve the compact-core and 
extended-dust 
in NGC\,3656; however, the fluxes of these components could be
of comparable magnitude, 
originate from very distinct emission mechanisms, and thus impact
their submm SED
analysis, as is the case in the 
easier to spatially resolve, nearest 
merger-remnant elliptical Centaurus\,A (Leeuw et al. 2002). 

The above results imply: (1) 350\,$\mu$m observations are not just
tracing cold dust, but also star-formation heating sources; (2)
longer (and achievable) 350\,$\mu$m SHARC\,II integrations are required
to detect the cold dust emission at lower flux 
contours
more closely associated with the CO and less heated by star-formation sites; (3) 
future mapping
sampling {\sl colder} and {\sl
  warmer} dust respectively using 
SCUBA\,II/JCMT at 850\,$\mu$m and HAWC/SOFIA at 88 and 155\,$\mu$m or
MIPS/$Spitzer$ at 24 and 70\,$\mu$m  
will be useful in 
de-convolving the star-formation heating effects from
the dust distribution; (4) follow-up interferometry at submm, mm or
high-frequency radio respectively with instruments such as SMA, CARMA, 
or VLA could also 
resolve the components of the cold dust,
star-forming, and compact core (if it is distinct) at spatial
resolutions comparable to the $HST$ maps.
Though our SHARC\,II observations are
preliminary, the 
results already provide the best spatially resolved
FIR-submm extent of this galaxy
and of any elliptical 
$>40$\,Mpc to
date, showing that dust of 29\,K, 
probably heated by star-formation, 
exists out to at least $\sim
1.8$\,kpc in NGC\,3656. 

\begin{acknowledgments}
This work was supported in part by NSF grant number AST 0204886.
LLL's participation at
the Terschelling meeting was partially funded by 
international travel grants from the American and
Royal Astronomical Societies. 
\end{acknowledgments}

\end{document}